\documentclass{article}
\usepackage{varioref}
\usepackage{amssymb}
\usepackage{color}

\begin{document}


\title{\bf  Massive gravity solution of Black Holes
and Entropy Bounds  }
\author{Hamed Hadi\thanks{email: hamedhadi1388@gmail.com} 
\\{\small Department of Physics, Azarbaijan Shahid Madani University, Tabriz}\\
\\Amin Rezaei Akbarieh \thanks{email: am.rezaei@tabrizu.ac.ir}
\\{\small Faculty of physics, University of Tabriz, Tabriz, Iran}\\
\\Pouneh Safarzadeh Ilkhchi\thanks{Pouneh.safarzadeh.1992@gmail.com}
\\{\small Faculty of physics, University of Tabriz, Tabriz, Iran}}
\date{\today} \maketitle
\begin{abstract}
The dRGT massive gravity represent a comprehensive theory which properly describes massive graviton field. Latterly, the exact spherical solutions are identified for the black hole in the dRGT massive gravity theory. In this paper, we derive Bousso's D-bound entropy for the black hole solutions of dRGT massive gravity. By an entropic consideration which provides a criterion, it is demonstrated that the relation between the D-bound and Bekenstein entropy bound imposes some constraints on the structure parameters of black hole solutions in dRGT massive gravity.
\end{abstract}
\section{Introduction}
Although more than a century has passed since Einstein came to dominate the theory of general relativity, recent observations show that we need to move toward a reform of the theory of general relativity and think more seriously about modified gravitational theories. For example, the nature of dark matter has remained unknown until now, and many of our conflicts have been between different theories \cite{Clifton:2011jh}. Supernova observations and other observations also led to the acceptance of the accelerating expansion of the universe as a fact among cosmologists \cite{SupernovaSearchTeam:1998fmf}. Many attempts have been made over the decades to accurately describe dark energy theoretically, and very important work has been published in this field \cite{Copeland:2007zz}. However, one of the most important theories for dark energy can be considered the theory of massive gravity. In the last decade, the massive gravity model has received much attention and its various aspects have been studied \cite{Hinterbichler:2011tt}.  In this model, graviton is assumed to be a massive spin-2 particle, and this feature of the gravitational force-mediated particle causes a fundamental change in the results and predictions in cosmology \cite{deRham:2014zqa}. Therefore, massive gravity theory addresses the problems in the standard cosmological model, such as describing the current accelerated expansion and the small cosmological constant compared to the theoretical value, with a different approach \cite{deRham:2010kj}.\\
 Many models have been developed to describe massive graviton and its interactions, but the main question is how to make a stable and consistent theory (theoretically and observationally). The development of a new model for describing the physics of a spin-2 massive particle started in 1939 by Fierz and Pauli, They expressed the linear action of a massive particle with spin-2 in a flat space \cite{Fierz:1939ix}. A few years later, Vainshtein provided an answer to the problem of vDVZ discontinuity that van Dam, Veltman and Zakharov  found in the action of Fierz and Pauli at the limit of $m\rightarrow 0$ \cite{Zakharov:1970cc}. In fact, he argued that Fierz and Pauli's theory should be nonlinear to avoid problems with the vDVZ discontinuity. But in 1979, Boulware and Deser claimed that Fierz and Pauli's nonlinear theory suffered from an instability, and it was referred to as the Boulware and Deser ghost \cite{Boulware:1972yco}. Finally, in 2011, Claudia de Rham, Gabadadze, and Tolley proposed a nonlinear theory of massive gravity   without ghost in a  specified decoupling limit that called dRGT massive gravity \cite{deRham:2010kj}. Since the dRGT massive gravity theory is unstable in flat-homogeneous FLRW space-time, the motivation to propose extended models increased\cite{Akbarieh:2022ovn}.\\
 In the literature, solutions of the black holes have been extensively studied in the context of massive gravity. One can find the Schwarzschild and Reissner–Nordström de-Sitter black hole metrics are considered as a precise solution to dRGT massive gravity theory \cite{Nieuwenhuizen:2011sq,Li:2016fbf}. A new type of solutions that describe charged black holes in massive gravity is presented in the paper \cite{Babichev:2014fka}, that exhibits an extra symmetry. Charged black holes in the Gauss–Bonnet gravity framework are another type of massive gravity black hole. A brief review of these solutions can be found in articles \cite{Volkov:2014ooa,Babichev:2015xha}. The dRGT massive gravity theory is one of the modified gravity theories used to explain the current acceleration in the expansion of the universe \cite{Tannukij:2017jtn}. The authors of this paper considered static spherical symmetric solutions in dRGT theory and in addition to the Schwarzschild and Reisner-Nordstrom solutions, they have introduced five other solutions. One of the solutions of the Einstein field equation in cylindrical symmetry is the black string \cite{Tannukij:2017jtn}. In the presence of quantum effects, black strings can emit a perturbation called Hawking radiation. Hawking radiation at spatial infinity differs from radiation at the source, due the gray body factor. Examining the exact boundaries of gray body factor from dRGT black string shows that plays an important role in the potential shape \cite{Boonserm:2019mon}. Therefore, studying a similar model gives the solution of a rotating black string in the presence of massive gravity \cite{Ghosh:2019eoo}.\\
 In the paper \cite{Rosen:2017dvn} with a static spherically-symmetric ansatz two types of black hole solutions in dRGT massive gravity are obtained, (1) Schwarzschild solutions that do not show Yukawa suppression at long distances and (2) Solutions where metrics are diagonal and have singularity on the horizon. In this work, the possibility of black hole solutions that can accommodate both the non-singular  horizon and Yukawa asymptotics is investigated. By adopting a time-dependent ansatz perturbative analytic solutions with non-single horizons are obtained, so in finite graviton mass they are time-dependent, but the apparent horizon location of the black hole is not necessarily time-dependent.
 By examining the spherical symmetric solutions of the dRGT massive gravity black hole in the presence of the cosmological constant, it can be shown that there are two types of solutions. In one of these solutions, the effect of massive potential appears as an effective cosmological constant \cite{Jafari:2017ypl}.

 In this paper, we consider the black hole solution in dRGT massive gravity under the scrutiny of D-bound \cite{Bousso:2000md} and Bekenstein bound \cite{{Bekenstein:1973ur},{Bekenstein:1980jp},{Bekenstein:1984vm}}. In doing so, we obtain these entropy bounds for black hole solution in dGRT gravity for different possibilities. Then we discuss the dilute system for this solution by entropic consideration. This consideration is a criterion which claims that the D-bound and Bekenstein bound must lead to the same physical entropy bound. Since two of these bonds are direct result of generalized second law of thermodynamics (GSL) over a certain mass, one demands their identification \cite{Bousso:2000md}. Requirements and consequences of this identification force some constraints over parameters of black hole solution in dRGT gravity.The solution of dRGT massive gravity by considering the equation of motions leads to some structure parameters which gives different kind of solutions such as black hole in cosmological-like fields. Considering entropic bounds and demanding identification of them imposes some constraints over these structure parameters and cosmological-like fields

 In the following we review briefly these bounds. Bekenstein bound puts an upper entropy bound over physical system which is isolated and stable thermodynamically within an asymptotically flat space as follows

 \begin{equation}
 	 S_{m}\leq 2\pi RE
 \end{equation}

 where $E$ is the total energy and $R$ is the radius of the system. Bekenstein bound has been investigated in empirical form which can be found in  $\cite{{Bekenstein:1984vm},{Schiffer:1989et},{Page:2000up},{Bekenstein:2000sw},{Wald:1999vt}}$. In addition its logical form which is based on generalized second law of thermodynamics (GSL) and the Geroch process, has been studied in $\cite{{Bekenstein:1980jp},{Bekenstein:1972tm},{Bekenstein:1974ax}}$. One can also find the quantum effects over this bound in $\cite{{Unruh:1982ic},{Unruh:1983ir},{Pelath:1999xt}}$.

 D-bound is other entropy bound which has been introduced in $\cite{Bousso:2000md}$. For a gravitational system one can consider the following gedanken experiment as follows. Suppose a matter system (black hole) with a cosmological horizon in the universe. For an observer  who is near the black hole, the entropy of the system is $S_{m}+S_{h}$ that $S_{m}$ and $S_{h}=A_{c}/4$ are the entropies of matter system and cosmological horizon, respectively. After recognizing the whole entropy of the system, the observer leave her location to far away from black hole until the matter system (black hole) be in her apparent cosmological horizon. In a thermodynamical process she will be witness of crossing the black hole from cosmological horizon. In this situation the final total entropy of the system for the observer is $S_{0}=A_{0}/4$ which $A_{0}$ is the area of the cosmological horizon when there is no matter system inside it. According to generalized second law of thermodynamic there is an upper bound over the matter system as follows
 \begin{equation}
 	S_{m}\leqslant \frac{1}{4}(A_{0}-A_{c})
 \end{equation}
This is D-bound over the matter system which has been introduced on de-Sitter space for the first time in $\cite{Bousso:2000md}$.

 The identification of D-bound and Bekenstein bound has been considered for the de-Sitter background and also identification is generalized for arbitrary dimensions $\cite{Bousso:2000md}$. Following the same method, the D-bound entropy for the various possible black
 hole solutions on a four dimensional brane, and its identification with the Bekenstein bound have been
 investigated in $\cite{Heydarzade:2017xbb}$. It is shown that there is a discrepancy for D-bound in higher dimensions in comparison with black hole solution in usual four dimensions. This difference is related to the extra loss of information due to the extra dimensions
 when black hole is crossing from the apparent horizon of the observer confined to the four dimensional brane. In addition McVitte solution of black hole which surrounded by cosmological fields has been considered in $\cite{Hadi:2019mtz}$. Demanding the identification of D-bound and Bekenstein bound leads to excluding quintessence and phantom background fields among cosmological constant. In $\cite{Hadi:2019lnm}$ again the identification of entropy bounds demands viability of cosmological constant field among other fields. For other example refer to $\cite{Hadi:2019qxn}$.

  It is important to note that since D-bound and Bekenstein bound are the direct result of GSL over a certain matter system, demanding the identification of both is logical physically. And also, Bekenstein bound is the maximum entropy bound over matter system then any deviation of it by D-bound is not physical or must have a physical interpretation such as extra energy or dimensions which is not inserted in Bekenstein bound.

 In this paper we will discuss all of the above considerations for black hole solution in dRGT massive gravity. The organization of the paper is as follows. The dRGT massive gravity and black hole solution is reviewed briefly in section 2. The Bekenstein bound and D-bound for different case of structure parameters of dRGT massive gravity black hole such as $(c=0 ~or~ \gamma = \zeta =0)$, $(\Lambda =0)$, $(\Lambda =0 , \zeta =0)$ and $(\Lambda \not= 0 , \gamma \not= 0, \zeta \not=0)$ are considered in sections 3, 4, 5 and 6 respectively. At the end we have a conclusion section.

\section{Black holes in dRGT massive gravity}
In this section, we review the dRGT massive gravity which is a nonlinear generalization solution of massive gravity which is free of BD ghost by incorporating higher order interaction terms into Lagrangian. One can indicate that the dRGT massive
gravity interacting with a scalar field with Einstein gravity. Therefore its action is Einstein-Hilbert action with suitable nonlinear term which is added to it and given as follows
\begin{equation}
S= \int d^{4}x\sqrt{-g}\frac{1}{2k^{2}}(R+m_{g}^{2}\mathcal{U}(g,\phi^{a}))
\end{equation}
where $R$ is the Ricci scalar and $\mathcal{U}$ is a potential for the graviton
which modifies the action with the parameter
$m_g$ as a interpretation of graviton mass. The action is written in units ($G=1$ thus, $k^2=8\pi$). The effective potential $\mathcal{U}$ in four-dimensional
spacetime can be written as follows
\begin{equation}
\mathcal{U}(g,\phi^2)=\mathcal{U}_{2}+\alpha_{3}\mathcal{U}_{3}+\alpha_{4}\mathcal{U}_{4}
\end{equation}
 where $\alpha_3$ and $\alpha_4$ are dimensionless free parameters of the theory. The potentials and their dependence to metric $g$ and scalar field
$\phi^{a}$ are given by
\begin{equation}
\mathcal{U}_{2}\equiv \left[ \mathcal{K} \right]^{2}-\left[ \mathcal{K}^{2}\right]
\end{equation}
\begin{equation}
\mathcal{U}_{3}\equiv \left[ \mathcal{K} \right]^{3}-3\left[ \mathcal{K} \right]\left[ \mathcal{K}^{2} \right]+2\left[ \mathcal{K}^{3} \right]
\end{equation}
\begin{equation}
\mathcal{U}\equiv \left[ \mathcal{K} \right]^{4}-6\left[ \mathcal{K} \right]^{2}\left[ \mathcal{K}^{2} \right]+8\left[ \mathcal{K} \right]\left[ \mathcal{K}^{3} \right]+3\left[ \mathcal{K}^{2} \right]^{2}-6\left[ \mathcal{K}^{4} \right]
\end{equation}
where
\begin{equation}
\mathcal{U}^{\mu}_{\nu}=\delta^{\mu}_{\nu}-\sqrt{g^{\mu\sigma}f_{ab}\partial_{\sigma}\phi^{a}\partial_{\nu}\phi^{b}}
\end{equation}
and $f_{ab}$ is a reference metric and $[\mathcal{K}]=\mathcal{K}^{\mu}_{\mu}$
is a trace. The $\phi^{a}$ are introduced as scalar fields to restore the
general covariance of the theory which are so-called St\"uckelberg scalar.

Now we choose a unitary gauge $\phi^{a}=x^{\mu}\delta^{a}_{\mu}$ . In this
gauge, the five degree of graviton is describe by the metric $g_{\mu \nu}$.
The St\"uckelberg scalar transform by the coordinate transformation. Due to the choosing of the unitary gauge one can break the gauge condition and
does an addition change in the st\"uckelberg scalers. Also one can redefine
the two parameters $\alpha_{3}$ and $\alpha_{4}$ according to new parameters as follows
\begin{eqnarray}
\alpha_{3}&=&\frac{\alpha-1}{3},\nonumber\\
 \alpha_{4}&=& \frac{\beta}{4}+\frac{1-\alpha}{12}
\end{eqnarray}
The modified field equations are obtained by varying the action with respect
to $g_{\mu \nu}$,
\begin{equation}\label{xxx}
G_{\mu \nu}+ m_{g}^{2}X_{\mu \nu}=0
\end{equation}
where
\begin{eqnarray}
X_{\mu \nu}&=&\mathcal{K}_{\mu \nu}-\mathcal{K}g_{\mu \nu}\nonumber\\
&-&(\mathcal{K}_{\mu \nu}^{2}-\mathcal{K}\mathcal{K}_{\mu \nu}+ \frac{[\mathcal{K}]^{2}-[\mathcal{K}^{2}]}{2}g_{\mu
\nu})\nonumber\\
&+&3\beta(\mathcal{K}_{\mu \nu}^{3}-\mathcal{K}\mathcal{K}_{\mu \nu}^{2}+
\frac{1}{2}\mathcal{K}_{\mu \nu}([\mathcal{K}]^{2}-[\mathcal{K}^{2}]))\nonumber\\
&-&\frac{1}{6}g_{\mu \nu}([\mathcal{K}]^{3}-3[\mathcal{K}][\mathcal{K}^{2}]+2[\mathcal{K}^{3}]).
\end{eqnarray}
$X_{\mu \nu}$ is the effective energy-momentum tensor.
In addition Bianchi identities give a constraint as follows
\begin{equation}
\bigtriangledown^{\mu}X_{\mu \nu}=0
\end{equation}
Notice that since $\bigtriangledown^{\mu}$ is covariant derivative and is compatible with $g_{\mu \nu}$, one can use $\alpha$ and $\beta$ instead of parameters $\alpha_{3}$ and $\alpha_{4}$.
\subsection{Black hole solution in dRGT massive gravity}
In this subsection we review static and spherically symmetric
black hole solution of the modified Einstein equations (\ref{xxx}) with line
element as follows

\begin{equation}
ds^{2}=-n(r)dt^{2}+2d(r)dtdr+\frac{dr^{2}}{f(r)}+h(r)^{2}d\Omega^{2}
\end{equation}
The solution can be classified into $d(r)=0$ or $h(r)=h_{0}r$ where $h_{0}$
is a constant in term of parameters $\alpha$ and $\beta$ \cite{{Koyama:2011yg},{Koyama:2011xz},{Sbisa:2012zk}}.  Because of simplicity we are interested in diagonal branch  with $d(r)=0$ and in order to have a black hole solution one can choose $h(r)=r$ leading to
\begin{equation}\label{metricsym}
ds^{2}=-n(r)dt^{2}+\frac{dr^{2}}{f(r)}+r^{2}d\Omega
\end{equation}
Here we choose the particular class solution of dRGT massive gravity with the symmetries that we are interested in them. As we know these symmetries restricts the degree of the freedom of the solutions and also it has important effect on the stability theory at times. Also note that most of the solution are asymptotically de Sitter or anti-de Sitter which confirms that at large scale the theory should recover the cosmological solution and in it the graviton mass will play the role of cosmological constant in order to explain the late time acceleration of the Universe.

Now solving equation (\ref{xxx})by the metric (\ref{metricsym}) one can obtain the solution for $f(r)$ and $n(r)$ as follows

 \begin{equation}\label{metric}
f(r)=n(r)
= 1 - \frac{2M}{r}+ \frac{\Lambda}{3} r^2 +\gamma r +\zeta,
\end{equation}
where
\begin{equation}
\Lambda=3m_{g}^{2}(1+\alpha+\beta)
\end{equation}
\begin{equation}
\gamma=-cm_{g}^{2}(1+2\alpha+3\beta)
\end{equation}
\begin{equation}
\zeta=c^{2}m_{g}^{2}(\alpha+3\beta)
\end{equation}
For details of solution refer to $\cite{Ghosh:2015cva}$. Here $M$ is an integration constant which is related to the mass of the black hole. There is a cosmological constant term $\Lambda$ which depends on the graviton mass $m_{g}$. The graviton mass has an  interpretation for the accelerated expansion of the Universe which comes consistently from the theory itself and one does not need to insert ad hoc cosmological constant $\Lambda$. Here the  Schwarzschild solution can be obtain by $m_{g}$ as one expected. The Schwarzschild-de Sitter and Schwarzschild-Anti de Sitter are derived by putting $c=0$ with the conditions $(1+\alpha+\beta)<0$ and $(1+\alpha+\beta)>0$, respectively.

\section{The Entropy Bounds for the Case of $c=0$ or ($\gamma=0$, $\zeta=0$).} This solution corresponds to the metric function
\begin{equation}\label{ff}
f(r)
= 1 - \frac{2M}{r}+\frac{\Lambda}{3} r^2,
\end{equation}
This solution is the same as Schwarzschild-de Sitter and  it
is de-Sitter with a positive cosmological constant $\frac{\Lambda}{3}$ asymptotically.
Such a solution is introduced in \cite{Bousso:2000md} as the example for the D-bound on matter entropy
in de Sitter space. One can infer some
physical results by considering the D-bound and its relationship with Bekenstein entropy bound.

Now we construct D-bound for this solution and consider its relationship with Bekenstein bound in order to drive some physical results for black holes
in  dRGT and restrict this solution by putting constraints over the parameters.

In order to construct D-bound for this case we need the initial entropy of the system which is sum of entropies of matter system and cosmological horizon as follows
\begin{equation}\label{entropyi}
S= S_m+\frac{A_c}{4},
\end{equation}
The final entropy of the system as we mentioned by details  in section (1) is when there is no matter system (black hole) in the bulk is $S_{0}=A_{0}/4$. Now according to GSL we have
\begin{equation}\label{dbound}
S_m\leqslant \frac{1}{4}(A_0-A_c).
\end{equation}
In order to derive this entropy bound we need to find the location of cosmological and black hole horizons. Setting $f(r)=0$ in equation (\ref{ff}) gives the  locations of the black hole horizon $r_b$ and cosmological horizon  $r_c$ in the presence of
black hole.

Now, we account the entropy of matter which enclosed by the cosmological horizon as follows,
\begin{equation}\label{sm}
S_m=\pi (r_c^2+r_b^2),
\end{equation}
 which is the sum of entropy of the black hole and cosmological horizon. The final entropy of the system when there is no any black hole in the bulk $(M=0)$ is given by
\begin{equation}
S_0=\pi r_0^2.
\end{equation}
where $r_{0}>r_{c}$ is the radius of cosmological horizon in pure de Sitter space. By solving the cubic equation (\ref{ff}), i.e f (r) = 0, and finding its positive
roots and putting them into (\ref{sm}), we can define relation (\ref{sm}) in the following form for small parameter $M$
\begin{equation}
S_m=\pi r_0^2(1-\frac {2M}{r_0})+ O(M^2).
\end{equation}
The energy of the system is not well-defined in
de Sitter space because of its unsuitable asymptotic region. However, Birkhoff's theorem implies that there exists some Schwarzschild-de
Sitter solution for a spherical system such that its metric is the same as the metric at large radii. One can define this large radius as cosmological horizon radius $r_{c}$. Therefore this black hole is called the "system's equivalent
black hole" and $r_{g}=2m$ is the gravitational radius of the system. In flat space, when there is no cosmological constant field, the radius $r_{g}$ reduces to Schwarzschild radius $2M$.

The mass parameter of the black hole $M$ can be derived by equation (\ref{ff}) and using $(r_{0})^{2}=-3/\Lambda$ which $\Lambda<0$ for Schwarzschild-de Sitter space. Therefore for mas parameter $M$ we have
\begin{equation}\label{3m}
2M=r_b(1-\frac{r_b^2}{r_0^2}).
\end{equation}
Now one can find the root $f(r_{c})=0$ for cosmological horizon and by inserting $2M$ from equation (\ref{3m}) the following equation is obtained
\begin{equation}\label{roo}
r_0^2=r_c^2(1+\frac{r_g}{r_c})+ O\left[(\frac{r_g}{r_c})^2\right].
\end{equation}
Note that we use the dilute system limit $(r_{g}<<r_{c})$ which admits limit of small "equivalent black holes" corresponding to light matter system. Therefore the D-bound for dilute system can be derived by using equation (\ref{roo}) and inequality (\ref{dbound}) is given by
\begin{equation}\label{bb}
S_m\leq \pi r_gr_c
\end{equation}
Now,  recall the Bekenstein bound for asymptotically non-flat spaces $\cite{Bousso:2000md}$, defined in terms of
gravitational radius $r_g=2m$ as
\begin{equation}\label{nn}
S_m\leq \pi r_gR,
\end{equation}
where $R=r_{c}$ is the radius of the system. Then, in Schwarzschild-de Sitter space for dRGT black hole solution of massive gravity with $(1+\alpha+\beta)<0$ Bekenstein bound and D-bound are the same in the limit of dilute system.

It is important to note that for the Schwarzschild-Anti de Sitter space $(1+\alpha+\beta)>0$ which correspond to positive cosmological constant there is no cosmological horizon $(r_{0})^{2}=-1/(m_{g}^{2}(1+\alpha +\beta))<0$. Therefore constructing D-bound and comparing it with Bekenstein bound in this case are not possible.
 \section{The Entropy Bounds for the Case of $\Lambda\simeq0$.}
The corresponding solution is given by
\begin{equation}
f(r)
= 1 - \frac{2M}{r}+ \gamma r + \zeta.
\end{equation}
This solution except for $\zeta$ term,  looks like to the Kiselev solution representing a
black hole surrounded by a quintessence field with the field structure parameter $\gamma<0$ $\cite{Kiselev:2002dx}$. Analogously we call this solution as Schwarzschild-quintessence-like black holes. In order to construct D-bound one needs to find the location of horizons. In doing so we find the positive roots of following equation
\begin{equation}\label{f}
	 1 - \frac{2M}{r}+ \gamma r + \zeta=0.
\end{equation}
Notice that for $M=0$ or equivalently in the absence of  black holes,  the above equation gives
 \begin{equation}\label{r0}
 r_0=\frac{1+\zeta}{-\gamma},
 \end{equation}
 which is the location of cosmological horizon when there is no back hole in the bulk and it is full of solely quintessence-like field. In addition, in this case for Schwarzschild-quintessence-like black hole we have $\gamma<0$ then $\zeta>-1$ in order to have positive amount for cosmological horizon radius $r_{0}$.

  In the presence of black hole, for $\gamma < 0 $  and $(1+\zeta)> 0$,  there are two   solutions of (\ref{f}) as

  \begin{equation}\label{rC}
r_{c}=\frac{-(1+\zeta )-\sqrt{(1+\zeta)^2 +8\gamma M}}{2\gamma },
\end{equation}
and \begin{equation}\label{rrB}
r_{b}=\frac{-(1+\zeta )+\sqrt{(1+\zeta)^2 +8\gamma M}}{2\gamma }
\end{equation}
It is important to note that the above roots the relations (\ref{rC}) $r_{c}$ and (\ref{rrB}) $r_{b}$ are positive for $\gamma<0$ and $1+\zeta>0$. Therefore in Schwarzschild-quintessence like solution there are roots for the location of horizons. Since we are interested in constructing D-bound, we do consider the Schwarzschild-quintessence like solution with parameters $\gamma<0$ and $1+\zeta>0$.

In doing to construct entropy bound we do as follows
 by rewriting $r_{c}^2$ as
\begin{equation}\label{rzeroo}
r_c^2=\frac{r_{c}(1+\zeta)-2M}{-\gamma}
\end{equation}
 The entropy related to cosmological horizon is $\pi r_{c}^{2}$. The initial entropy of the system (black hole plus cosmological horizon) is $S = S_{m}+A_{c}/4$. Here, $S_m$ is the entropy of the matter (black hole) inside the cosmological horizon and $A_c$ is the area of cosmological horizon in the presence of  black hole.  At the end of process of gedenken experiment for constructing D-bound when there is no black hole in the bulk, the
final entropy of the system will be $S_0=A_0/4$, using the fact that the quarter of the area of the
apparent cosmological horizon is Bekenstein-Hawking entropy. Here $A_0$ is the area of
horizon of empty cosmological space. Then, the  generalized second law ($S\leq S_{0}$) leading  to D-bound (\ref{dbound}) which is given by

\begin{equation}\label{Nbulk}
S_m\leqslant \pi\left( (\frac{1+\zeta}{\gamma})^{2}+\frac{r_{c}(1+\zeta)}{\gamma}-\frac{2M}{\gamma}\right).
\end{equation}
  This is D-bound in general form for this solution. Now we derive this bound for a dilute system $(r_{g}<< r_{c})$. In other words one can consider small black hole $(M\rightarrow 0)$ in dilute system. Recalling the approach we had in section (3), here the system equivalent black hole radius is $r_{g}=r_{b}$. Therefore in dilute limit the relation (\ref{rrB}) becomes
  \begin{equation}\label{diluterg}
  	r_{g}=r_{b}=-\frac{2M}{\gamma r_{0}} +O(M^{2})
  \end{equation}
  where we used the relation (\ref{r0}). Now we insert this limit into inequality (\ref{Nbulk}) and using the relation $r_{g}+r_{c}=r_0 $, we obtain

\begin{equation}
S_m\leqslant \pi (r_{0}r_{g}+r_{0}(r_{0}-r_{c})).
\end{equation}
Since $r_{0}>r_{c}$ the bound is positive. However in this case D-bound is not the same as Bekenstein bound (\ref{nn}) with $R=r_{c}$ the cosmological horizon as follows $S_{m}\leqslant \pi r_{g}r_{c}$. The Schwarzschild-quintessence like field in this model does not satisfy the identification of D-bound and Bekenstein bound.
\section{The Entropy Bounds for the Case of $\Lambda=0$ and  $\zeta=0$.}
The corresponding solution is given by

\begin{equation}
	f(r)
	=1-\frac{2M}{r}+\gamma r
\end{equation}

which is the Schwarzschild black hole in the quintessence field $\cite{Kiselev:2002dx}$ if the
quintessence structure parameter is $\gamma<0$.
This solution occurs for parameters $\alpha=-3/2$ and $\beta=1/2$. Now,  in order for our purpose for constructing D-bound we find the solution of the following equation,
\begin{equation}\label{fb}
	f(r) =1-  \frac{2M}{r}+\gamma r =0.
\end{equation}
which gives the location of the horizons.
For $M=0$ or equivalently in the absence of  black hole,  the above equation gives
\begin{equation}\label{fracc}
	r_0 =\frac{1}{-\gamma}
\end{equation}
representing the radius of cosmological horizon of an empty of matter system and is positive for $\gamma <0$ . In the presence of black hole, there are two   solutions of (\ref{fb}) as
\begin{equation}\label{rB}
	r_{b}=\frac{-1+\sqrt{1+8\gamma M}}{2\gamma },
\end{equation}
and \begin{equation}\label{rCD}
	r_{c}=\frac{-1-\sqrt{1 +8\gamma M}}{2\gamma },
\end{equation}
representing black hole horizon and cosmological horizon in the presence of black hole, respectively. Notice that $\gamma <0$ and $1+8\gamma M >0$ and also $1-\sqrt{1+8\gamma M}>0$. Hence we have $-1< 8\gamma M <0$ which leads to $0< \sqrt{1+8\gamma M} <1$. Therefore without any contradiction we have two horizons for this solution.
Here, it is useful to rewrite (\ref{rCD}) as
\begin{equation}\label{fractal}
	r_c ^2=\frac{-1}{\gamma}(r_c -2M).
\end{equation}
We know from previous sections that $r_{g}=r_{b}$ is the gravitational radius of equivalent black hole system. Then, by using $ r_{g} + r_{c}= r_{0}$ and the  generalized second law ($S\leq S_{0}$) leading  to
D-bound (\ref{dbound}) we find
\begin{eqnarray}\label{cosmology}
	S_m&\leq&\pi r_0 ^2 -\pi r_c ^2\nonumber\\
	&=&\pi((r_c+r_g)^2 -r_0(r_c -2M))\nonumber\\
\end{eqnarray}
where in the second line we use equations (\ref{fracc}) and (\ref{fractal}). In the limit of dilute system $(r_{g}<<r_{c})$ the inequality (\ref{cosmology}) becomes
\begin{equation}
	S_{m} \leq 2\pi r_{g}r_{c}
\end{equation}
This is D-bound for Schwarzschild-quintessence like solution for dilute system limit. The Bekenstein bound (\ref{nn}) for $R=r_{c}$ is tighter than D-bound in this case.
\section{The Entropy Bounds for the Case of $\Lambda\neq 0$ , $\gamma\neq0$
and $\zeta\neq0$}
The corresponding solution is given by
\begin{equation}\label{aba}
f(r)=1 - \frac{2M}{r}+ \frac{\Lambda}{3} r^2 +\gamma r +\zeta
\end{equation}
In the absence of black hole, i.e $M=0$,  the above equation gives
 \begin{equation}\label{c}
\frac{\Lambda}{3}r_{0}^{2}+\gamma r_{0}=-(1+\zeta),
 \end{equation}
where $r_0$ is the radius of the cosmological horizon in the absence of any matter system  (black hole).In order to have positive roots for $r_{0}$ we have 
\begin{equation}\label{positive}
	r_{0}=\frac{-\gamma +\sqrt{\gamma^{2}-\frac{4}{3}\Lambda(1+\zeta)}}{\frac{2\Lambda}{3}}
\end{equation} 
where  $\gamma>0$ and $\Lambda(1+\zeta)<0$. Also we have $\gamma>\sqrt{\frac{4}{3}\Lambda(1+\zeta)}$. It is obvious that this root leads to a  contradiction for $\Lambda>0$ and there is no any root for $r_{0}$ here. However, for $\gamma>0$, $\Lambda<0$ and $1+\zeta<0$ there is a positive root here. On the other hand for $\gamma<0$, $\Lambda>0$ and $1+\zeta>0$ there is also a positive root for $r_{0}$. 

The other solution is as follows
\begin{equation}\label{negative}
r_{0}=\frac{-\gamma -\sqrt{\gamma^{2}-\frac{4}{3}\Lambda(1+\zeta)}}{\frac{2\Lambda}{3}}
\end{equation}
where for $\gamma>0$, $\Lambda<0$ and $1+\zeta<0$ we have a positive root for $r_{0}$ here.
On the other hand for $\gamma<0$ and $\Lambda(1+\zeta)<0$ are imposed in order to have a positive root for $r_{0}$ and also we have $0>\gamma>-\sqrt{\frac{4}{3}\Lambda(1+\zeta)}$. Since we are considering D-bound, it is necessary to have a positive solution for $r_{0}$ and it imposes the above constraints over the structure parameters of this solution. As we see this solution leads to contradiction. However, for $\gamma<0$, $\Lambda>0$ and $\Lambda(1+\zeta)>0$ there is a positive root for $r_{0}$.  Therefore there is two categories of restrictions for having positive roots for cosmological horizon $r_{0}$ in the absence of black hole as follows
\begin{itemize}
		\item $\gamma <0$ , $\Lambda>0$ and $1+\zeta>0$
		\item $\gamma >0$ , $\Lambda<0$ and $1+\zeta<0$
\end{itemize}
It is important to note that for constructing D-bound these constraints are necessary over structure parameters of dRGT massive gravity. To construct D-bound we need to find the solutions for cosmological horizon in the presence of black hole. Therefore we do as follows

 Using  equation (\ref{c}), one can rewrite  (\ref{aba}) as
 \begin{equation}\label{sigma}
f(r)=-\frac{\Lambda}{3}r_{0}^{2}-\gamma r_{0} - \frac{2M}{r}+ \frac{\Lambda}{3} r^{2} +\gamma r.
 \end{equation}
 Now,  by  solving $f(r)=0$, we can find two roots $r_c$ and $r_b$ as the cosmological horizon in the presence
 of matter system (black hole) and the black hole horizon as
 \begin{equation}\label{hfh}
-\frac{\Lambda}{3}r_{0}^{2}-\gamma r_{0} - \frac{2M}{r_{c}}+ \frac{\Lambda}{3} r^{2}_{c} +\gamma r_{c}=0
 \end{equation}
 and
  \begin{equation}\label{gg}
-\frac{\Lambda}{3}r_{0}^{2}-\gamma r_{0} - \frac{2M}{r_{b}}+ \frac{\Lambda}{3} r^{2}_{b} +\gamma r_{b}=0
 \end{equation}
  using (\ref{hfh}),  we have
\begin{equation}\label{dVt1}
r_{0}^{2}=r_{c}^{2}+\frac{3\gamma}{\Lambda}(r_{c}-r_{0})-\frac{6M}{r_{c}\Lambda}
\end{equation}
In order to derive D-bound we use (\ref{dVt1}) in (\ref{dbound}) which leads to
\begin{equation}\label{ddbouund}
	S_{m} \leq \pi (\frac{3\gamma}{\Lambda}(r_{c}-r_{0})-\frac{6M}{r_{c}\Lambda})
\end{equation}
Since $r_{0}$ is the radius of cosmological horizon in the absence of mass $M$ which provides an attractive force, we can infer that $r_{0}>r_{c}$. Therefore the above D-bound (\ref{ddbouund}) in order to have positive amount must put some constraints over structure parameters. We know from positive roots of $r_{0}$ that $\gamma \Lambda <0$, then the first term of D-bound $(\ref{ddbouund})$ is positive and in order to have positive entropy bound we must have $\frac{3\gamma}{\Lambda}(r_{c}-r_{0})>\frac{6M}{r_{c}\Lambda}$. Therefore we can both positive and negative amount for $\Lambda$. However in the dilute system as we show in following, $\Lambda$ must be negative. 

In the limit of dilute system when black hole mass is negligible then from relation (\ref{hfh}) one can infer that $r_{c} \approx r_{0}$. Therefore according to inequity (\ref{ddbouund}) results in
\begin{equation}
	S_{m} \leq -\pi \frac{6M}{r_{c}\Lambda}
\end{equation}
which is D-bound for dilute limit and in order to have positive bound we must have $\Lambda <0$. Notice that we maintain the term $\frac{6M}{r_{c}\Lambda}$ because of small amount of $\Lambda$ .

Like as previous cases, Bekenstein bound is given by $S_{m}<\pi r_{g}r_{c}$ with cosmological horizon radius $r_{c}$ as radius of system. Now we derive this bound in light system (dilute limit) as following. First we subtract relations (\ref{hfh}) and (\ref{gg}) from each others which is given by
\begin{equation}\label{subtract}
	\frac{2M}{r_{c}r_{g}}+\frac{\Lambda}{3}(r_{c}+r_{g})+\gamma =0
\end{equation}
Notice that we again used $r_{g}=r_{b}$ as gravitational radius. Now we obtain $r_{g}r_{c}$ in limit of dilute system $(r_{g}<< r_{c})$ by using equation (\ref{subtract}) as follows
\begin{equation}
	r_{g}r_{c}=-\frac{6M}{\Lambda r_{c}} +\frac{18\gamma M}{\Lambda^{2} r_{c}^{2}}
\end{equation}
Therefore the Bekenstein bound is given by
\begin{equation}
	S_{m} \leq -\pi \frac{6M}{\Lambda r_{c}} +\pi \frac{18\gamma M}{\Lambda^{2} r_{c}^{2}}.
\end{equation}
If we ignore the $\gamma$ term D-bound and Bekenstein bound is identified. Any term expect cosmological constant like term results in deviation of D-bound from Bekenstein bound in dilute system of black hole solution in dRGT massive gravity. Again in the limit of dilute system the Bekenstein bound is true for negative amount of $\Lambda$. 

At the end of this section we want to answer the important question. What happens with respect to D-bound and Bekenstein bound on the matter entropy when the black hole and cosmological horizons coincide? In this case there is a Bekenstein bound over the matter system and its upper bound is proportional to energy of the system and radius of cosmological or black hole horizon. However, defining D-bound, in this case, is not possible, because as we explained in the introduction section, to construct D-bound we need a cosmological horizon which is not coincide with black hole's horizon.  

\section{Conclusion}
We have considered black hole solution in dRGT massive gravity by respecting an entropic consideration. The criterion for this consideration is the identification of D-bound and Bekenstein bound which are the direct result of generalized second law of thermodynamics putting an upper bound over a certain matter system. Since D-bound and Bekenstein bound are the direct result of GSL over a certain matter system, demanding the identification of both is logical physically. We indicated for dRGT black hole solutions that this identification occurs solely for cosmological-constant-like fields and other terms with different structure parameters have a deviation from these bounds identification. Therefore this etropic consideration excludes cosmological and exotic fields expect the cosmological-constant-like fields and asserts that this is the only viable cosmological field among others.

It is concluded that for the case when structure parameters are $(c=0)$ or $(\gamma = \zeta =0)$, the D-bound and Bekenstein bound are identified which assert that Schwarzschild-de Sitter
space for dRGT black hole solution of massive gravity behaving like as cosmological constant-like field is a physical field and provide entropic consideration. This solution in the case of Schwarzschild-Anti de Sitter is not consistent with entropic criterion. The dRGT black hole solution for $\Lambda=0, \gamma <0, 1+\zeta>0$ has two horizons and the comparison of D-bound and Bekenstein bound excludes this odd cosmological field. The Schwarzschild-quintessence like solution of dRGT black hole $(\Lambda=\zeta=0 ,\gamma <0 )$ leads to tighter Bekenstein bound than D-bound then identification of these two bounds does not occur here. The Entropy Bounds for the Case of $\Lambda\neq 0$ , $\gamma\neq0$
and $\zeta\neq0$ for dilute system and by ignoring $\gamma$ are identified. As a conclusion it is claimed that any
term expect cosmological constant-like field results in deviation of D-bound
from Bekenstein bound in dilute system of black hole solution in dRGT massive gravity. 
\\
\textbf{Data Availability Statement}: No Data associated in the manuscript.



\begin{thebibliography}{99}
\bibitem{Clifton:2011jh}
T.~Clifton, P.~G.~Ferreira, A.~Padilla and C.~Skordis,
Phys. Rept. \textbf{513} (2012), 1-189
doi:10.1016/j.physrep.2012.01.001
[arXiv:1106.2476 [astro-ph.CO]].

\bibitem{SupernovaSearchTeam:1998fmf}
A.~G.~Riess \textit{et al.} [Supernova Search Team],
Astron. J. \textbf{116} (1998), 1009-1038
doi:10.1086/300499
[arXiv:astro-ph/9805201 [astro-ph]].

\bibitem{Copeland:2007zz}
E.~J.~Copeland,

AIP Conf. Proc. \textbf{957} (2007) no.1, 21-29
doi:10.1063/1.2823765

\bibitem{Hinterbichler:2011tt}
K.~Hinterbichler,

Rev. Mod. Phys. \textbf{84} (2012), 671-710
doi:10.1103/RevModPhys.84.671
[arXiv:1105.3735 [hep-th]].

\bibitem{deRham:2014zqa}
C.~de Rham,
Living Rev. Rel. \textbf{17} (2014), 7
doi:10.12942/lrr-2014-7
[arXiv:1401.4173 [hep-th]].


\bibitem{deRham:2010kj}
C.~de Rham, G.~Gabadadze and A.~J.~Tolley,
Phys. Rev. Lett. \textbf{106} (2011), 231101
doi:10.1103/PhysRevLett.106.231101
[arXiv:1011.1232 [hep-th]].


\bibitem{Fierz:1939ix}
M.~Fierz and W.~Pauli,
Proc. Roy. Soc. Lond. A \textbf{173} (1939), 211-232
doi:10.1098/rspa.1939.0140


\bibitem{Zakharov:1970cc}
V.~I.~Zakharov,
JETP Lett. \textbf{12} (1970), 312



\bibitem{Boulware:1972yco}
D.~G.~Boulware and S.~Deser,

Phys. Rev. D \textbf{6} (1972), 3368-3382
doi:10.1103/PhysRevD.6.3368



\bibitem{Akbarieh:2022ovn}
A.~R.~Akbarieh, S.~Kazempour and L.~Shao,
Phys. Rev. D \textbf{105} (2022) no.2, 023501
doi:10.1103/PhysRevD.105.023501
[arXiv:2203.00901 [gr-qc]].



\bibitem{Nieuwenhuizen:2011sq}
T.~M.~Nieuwenhuizen,
Phys. Rev. D \textbf{84} (2011), 024038
doi:10.1103/PhysRevD.84.024038
[arXiv:1103.5912 [gr-qc]].


\bibitem{Li:2016fbf}
P.~Li, X.~z.~Li and P.~Xi,
Phys. Rev. D \textbf{93} (2016) no.6, 064040
doi:10.1103/PhysRevD.93.064040
[arXiv:1603.06039 [gr-qc]].



\bibitem{Babichev:2014fka}
E.~Babichev and A.~Fabbri,

JHEP \textbf{07} (2014), 016
doi:10.1007/JHEP07(2014)016
[arXiv:1405.0581 [gr-qc]].



\bibitem{Volkov:2014ooa}
M.~S.~Volkov,
Lect. Notes Phys. \textbf{892} (2015), 161-180
doi:10.1007/978-3-319-10070-8\_6
[arXiv:1405.1742 [hep-th]].


\bibitem{Babichev:2015xha}
E.~Babichev and R.~Brito,
Class. Quant. Grav. \textbf{32} (2015), 154001
doi:10.1088/0264-9381/32/15/154001
[arXiv:1503.07529 [gr-qc]].



\bibitem{Tannukij:2017jtn}
L.~Tannukij, P.~Wongjun and S.~G.~Ghosh,
Eur. Phys. J. C \textbf{77} (2017) no.12, 846
doi:10.1140/epjc/s10052-017-5426-0
[arXiv:1701.05332 [gr-qc]].



\bibitem{Boonserm:2019mon}
P.~Boonserm, T.~Ngampitipan and P.~Wongjun,
Eur. Phys. J. C \textbf{79} (2019) no.4, 330
doi:10.1140/epjc/s10052-019-6827-z
[arXiv:1902.05215 [gr-qc]].



\bibitem{Ghosh:2019eoo}
S.~G.~Ghosh, R.~Kumar, L.~Tannukij and P.~Wongjun,
Phys. Rev. D \textbf{101} (2020) no.10, 104042
doi:10.1103/PhysRevD.101.104042
[arXiv:1903.08809 [gr-qc]].


\bibitem{Rosen:2017dvn}
R.~A.~Rosen,

JHEP \textbf{10} (2017), 206
doi:10.1007/JHEP10(2017)206
[arXiv:1702.06543 [hep-th]].




\bibitem{Jafari:2017ypl}
G.~Jafari, M.~R.~Setare and H.~R.~Bakhtiarizadeh,
Phys. Lett. B \textbf{773} (2017), 395-400
doi:10.1016/j.physletb.2017.08.057
[arXiv:1702.00189 [gr-qc]].


\bibitem{Bousso:2000md}
R.~Bousso,
JHEP \textbf{04} (2001), 035
doi:10.1088/1126-6708/2001/04/035
[arXiv:hep-th/0012052 [hep-th]].


\bibitem{Bekenstein:1973ur}
J.~D.~Bekenstein,
Phys. Rev. D \textbf{7} (1973), 2333-2346
doi:10.1103/PhysRevD.7.2333


\bibitem{Bekenstein:1980jp}
J.~D.~Bekenstein,
Phys. Rev. D \textbf{23} (1981), 287
doi:10.1103/PhysRevD.23.287


\bibitem{Bekenstein:1984vm}
J.~D.~Bekenstein,
Phys. Rev. D \textbf{30} (1984), 1669-1679
doi:10.1103/PhysRevD.30.1669


\bibitem{Schiffer:1989et}
M.~Schiffer and J.~D.~Bekenstein,
Phys. Rev. D \textbf{39} (1989), 1109-1115
doi:10.1103/PhysRevD.39.1109


\bibitem{Page:2000up}
D.~N.~Page,
[arXiv:hep-th/0007237 [hep-th]].


\bibitem{Bekenstein:2000sw}
J.~D.~Bekenstein,
[arXiv:gr-qc/0006003 [gr-qc]].


\bibitem{Wald:1999vt}
R.~M.~Wald,
Living Rev. Rel. \textbf{4} (2001), 6
doi:10.12942/lrr-2001-6
[arXiv:gr-qc/9912119 [gr-qc]].


\bibitem{Bekenstein:1972tm}
J.~D.~Bekenstein,
Lett. Nuovo Cim. \textbf{4} (1972), 737-740
doi:10.1007/BF02757029


\bibitem{Bekenstein:1974ax}
J.~D.~Bekenstein,
Phys. Rev. D \textbf{9} (1974), 3292-3300
doi:10.1103/PhysRevD.9.3292


\bibitem{Unruh:1982ic}
W.~G.~Unruh and R.~M.~Wald,
Phys. Rev. D \textbf{25} (1982), 942-958
doi:10.1103/PhysRevD.25.942


\bibitem{Unruh:1983ir}
W.~G.~Unruh and R.~M.~Wald,
Phys. Rev. D \textbf{27} (1983), 2271-2276
doi:10.1103/PhysRevD.27.2271


\bibitem{Pelath:1999xt}
M.~A.~Pelath and R.~M.~Wald,
Phys. Rev. D \textbf{60} (1999), 104009
doi:10.1103/PhysRevD.60.104009
[arXiv:gr-qc/9901032 [gr-qc]].


\bibitem{Heydarzade:2017xbb}
Y.~Heydarzade, H.~Hadi, C.~Corda and F.~Darabi,
Phys. Lett. B \textbf{776} (2018), 457-463
doi:10.1016/j.physletb.2017.11.061
[arXiv:1706.04434 [gr-qc]].


\bibitem{Hadi:2019mtz}
H.~Hadi, Y.~Heydarzade, F.~Darabi and K.~Atazadeh,
Eur. Phys. J. Plus \textbf{135} (2020) no.7, 584
doi:10.1140/epjp/s13360-020-00601-7
[arXiv:1910.09980 [gr-qc]].


\bibitem{Hadi:2019lnm}
H.~Hadi, F.~Darabi, K.~Atazadeh and Y.~Heydarzade,
Eur. Phys. J. C \textbf{80} (2020) no.12, 1126
doi:10.1140/epjc/s10052-020-08699-w
[arXiv:1903.05119 [gr-qc]].


\bibitem{Hadi:2019qxn}
H.~Hadi, F.~Darabi and Y.~Heydarzade,
EPL \textbf{131} (2020) no.5, 59001
doi:10.1209/0295-5075/131/59001
[arXiv:1907.07143 [gr-qc]].


\bibitem{Koyama:2011yg}
K.~Koyama, G.~Niz and G.~Tasinato,
Phys. Rev. D \textbf{84} (2011), 064033
doi:10.1103/PhysRevD.84.064033
[arXiv:1104.2143 [hep-th]].


\bibitem{Koyama:2011xz}
K.~Koyama, G.~Niz and G.~Tasinato,
Phys. Rev. Lett. \textbf{107} (2011), 131101
doi:10.1103/PhysRevLett.107.131101
[arXiv:1103.4708 [hep-th]].


\bibitem{Sbisa:2012zk}
F.~Sbisa, G.~Niz, K.~Koyama and G.~Tasinato,
Phys. Rev. D \textbf{86} (2012), 024033
doi:10.1103/PhysRevD.86.024033
[arXiv:1204.1193 [hep-th]].


\bibitem{Ghosh:2015cva}
S.~G.~Ghosh, L.~Tannukij and P.~Wongjun,
Eur. Phys. J. C \textbf{76} (2016) no.3, 119
doi:10.1140/epjc/s10052-016-3943-x
[arXiv:1506.07119 [gr-qc]].


\bibitem{Kiselev:2002dx}
V.~V.~Kiselev,
Class. Quant. Grav. \textbf{20} (2003), 1187-1198
doi:10.1088/0264-9381/20/6/310
[arXiv:gr-qc/0210040 [gr-qc]].


\end{thebibliography}
\end{document}